\documentclass[twocolumn,showpacs,amsmath,amssymb,pra]{revtex4}
  
\usepackage{graphicx}
\usepackage{dcolumn}
\usepackage{bm}
\usepackage{bbm}
\usepackage{amsfonts}

\def\Z{{\mathbb{Z}}}

\newcommand{\ve}[1]{\mathbf{#1}}
\newcommand{\bra}[1]{\langle #1|}
\newcommand{\ket}[1]{|#1\rangle}

\newcommand{\bracket}[3]{\langle #1|#2|#3 \rangle}
\newcommand{\tr}[2]{{\,\rm tr_{#1}}{\lbrack #2 \rbrack}\,}
\def\ii{{\rm i}}

\begin{document}

\title{Generation of entanglement in regular systems}
\author{Marko \v Znidari\v c and Toma\v z Prosen}
\affiliation{Physics Department, Faculty of Mathematics and Physics, 
University of Ljubljana, Ljubljana, Slovenia}

\begin{abstract}
We study dynamical generation of entanglement
in bipartite quantum systems, characterized by {\em purity} 
(or linear entropy), and caused by the coupling between the two subsystems.
Explicit semiclassical theory of purity decay is derived for 
integrable classical dynamics of the uncoupled system, and for 
localized (general Gaussian wave-packet) initial states.
Purity decays as an algebraic function of {\em time} $\times $ 
{\em strength of perturbation}, independently of the Planck's constant.
\end{abstract}

\pacs{3.65.Ud, 3.67.Mn, 3.65.Sq} 

\date{\today}

\maketitle
Detailed understanding of {\em entanglement}, being one of the most distinct
features of the quantum world, is an issue of high importance, particularly 
in view of recent efforts to build quantum devices that will manipulate 
(pure states of) individual quantum systems.
The loss of control over the entanglement, e.g. 
decoherence, in such a device is one of the major obstacles that we 
have to overcome. 

In the present paper we are going to study dynamical generation of 
entanglement in {\em bipartite} systems. Initially {\em separable} pure state will 
get entangled due to the coupling between two subsystems. Here we consider 
systems where the uncoupled part of the Hamiltonian in both subsystems
generates regular (integrable) dynamics in the classical limit.
The motivation to study entanglement generation 
in systems with a regular uncoupled dynamics comes from the fact that such 
systems are quite common both in experiments and as theoretical models. 
For instance, if the uncoupled system consists of a number of uncoupled 
one {\em degree of freedom} (DOF) systems then it is integrable. Such 
is the case in various proposals for quantum computation, e.g. ion traps. 
Further, experimentally realizable Jaynes-Cummings model, where decoherence for 
cat states~\cite{Haroche:96} has actually been 
experimentally measured, is also an integrable system. Still further, 
a standard model of decoherence~\cite{Vernon:63} consists of an infinite 
number of harmonic oscillators. If the bath consists of a finite number of 
harmonic oscillators this falls under the domain of our theory. 
Recently~\cite{Haake} it has been pointed out that 
the decoherence for truly macroscopic superposition is so fast that the 
usual master equation approach is not valid anymore. On this very short 
``instantaneous'' time scale {\em any} system will effectively behave as 
a regular one (i.e. correlations do not decay yet). The semiclassical 
decay of purity has recently been discussed by Jacquod~\cite{Jacquod:04},
however his results in the regular domain do not agree with our findings.
See \cite{other} for some other related recent results.

Time evolution of the system will be governed by a Hamiltonian
\begin{equation}
H=H_0+\delta\,\cdot V ,\qquad H_0=H_{\rm A} \otimes \mathbbm{1}_{\rm B}+\mathbbm{1}_{\rm A} \otimes H_{\rm B},
\label{eq:H}
\end{equation}
where $H_0$ is an uncoupled part of the Hamiltonian and $V$ is the coupling 
between the two subsystems responsible for the generation of entanglement. 
The strength of this coupling is given by a dimensionless parameter $\delta$. 
We will use subscripts ``A'' and ``B'' to denote two subsystems. 
The state of the whole system at time $t$ is simply 
$\ket{\psi(t)} =U(t)\ket{\psi(0)}$, with a unitary propagator 
$U(t)=\exp{(-\ii H t/\hbar)}$. Let us define time-averaged coupling
\begin{equation}
\bar{V}=\lim_{T \to \infty}{\frac{1}{T} \int_0^T{\!\!\!dt\, V(t)}},
\label{eq:Vbar}
\end{equation}
where $V(t)$ is the coupling operator in the {\em interaction picture}, 
$V(t)=U_0^\dagger(t) V U_0(t)$, $U_0(t)=\exp{(-\ii H_0 t/ \hbar)}$, 
i.e. propagated with an {\em uncoupled} part of the Hamiltonian.
We shall assume a situation, typical for a regular $H_0$, where 
$\bar{V}$ is a non-trivial operator, different from zero or a multiple of 
identity~\cite{Freeze}. We wish to stress that perturbation $V$ will typically break the
integrability of $H$, and that our results reported below are not limited
to values of small $\delta$.

The entanglement between the two subsystems, for a pure state $\ket{\psi(t)}$,
is characterized by a purity
\begin{equation}
I(t)=\tr{A}{\rho_{\rm A}^2(t)},\qquad \rho_{\rm A}(t)=
\tr{B}{\rho(t)}
\label{eq:I}
\end{equation}
where $\rho(t):=\ket{\psi(t)}\bra{\psi(t)}$.
Iff purity $I(t)$ is less than $1$, the two subsystems are entangled, 
otherwise they are in a separable (product) state. Our 
{\em initial} state will always be a product one, 
$\ket{\psi(0)}=\ket{\psi_{\rm A}(0)}\otimes \ket{\psi_{\rm B}(0)}$, hence $I(0)=1$. 
The initial states 
$\ket{\psi_{\rm A,B}(0)}$ will be Gaussian wave packets. Time dependence of 
purity $I(t)$ will then tell us how fast the two subsystems get entangled due to the 
coupling $V$. 

Let us proceed with the calculation of purity decay $I(t)$.
We should observe that propagating the state backwards in time 
with a {\em separable} - {\em uncoupled} dynamics $U_0(t)$ 
does not change the value of purity, so
$\rho(t)$ in Eq.~(\ref{eq:I}) can be replaced by
\begin{equation}
\rho^{\rm M}(t)=M(t) \rho(0) M^\dagger(t),\qquad M(t)=U^\dagger_0(t) U(t),
\label{eq:M}
\end{equation}
where $M(t)$ is the {\em echo operator} used in the theory of fidelity 
decay~\cite{Prosen:02spin,Prosen:03}. The matrix $\rho^{\rm M}(t)$ represents the
evolution of our pure state in the interaction picture.
As just explained above, the purity (\ref{eq:I}) is equal to
\begin{equation}  
I(t)=
\tr{A}{\{ \rho^{\rm M}_{\rm A}(t) \}^2},\qquad \rho^{\rm M}_{\rm A}(t)=
\tr{B}{\rho^{\rm M}(t)}.
\label{eq:IM}
\end{equation}
An advantage of the representation (\ref{eq:IM}) over (\ref{eq:I}) is the fact that
the echo operator $M_\delta(t)$ is, unlike the forward evolution 
$U(t)$, close to an identity for
small $\delta$ so one may use perturbative, or asymptotic expansions in $\delta$.
We follow the approach of Ref.~\cite{Prosen:03} and use the 
Baker-Campbell-Hausdorff formula $e^{\delta V} e^{\delta W} = 
\exp(\delta(V+W) + \frac{1}{2}\delta^2[V,W]+\ldots)$
for continuous products, see e.g.~\cite{Birula}, 
to simplify the expression for the echo operator $M(t)$.
The lowest order term in the exponential is $\frac{\delta}{\hbar}\int{\! dt V(t)}$. 
For times larger than some classical averaging time $t_{\rm ave}$, on which $\bar{V}$ 
(\ref{eq:Vbar}) converges, this term can be rewritten as $\frac{\delta}{\hbar}
\bar{V} t$. The second order term in $\delta$ can be shown to grow with time no 
faster than $\delta^2 t/\hbar$, 
and by induction higher orders can be estimated to grow as 
$\sim\delta^r t^{r-1}/\hbar$. 
Therefore, provided only $\delta \ll 1$, higher orders in $\delta$ can be neglected 
and we end up with a very simple expression for the echo operator
\begin{equation}
M(t)={\rm e}^{-\ii \delta t \bar{V}/\hbar}.
\label{eq:Mvbar}
\end{equation}
So the echo operator can be interpreted as the propagator with an 
effective Hamiltonian $\delta \bar{V}$.
We proceed with a semiclassical evaluation of the purity, a procedure completely 
analogous to a similar calculation for the fidelity~\cite{Prosen:02JPA}.
We use the notation in which small Latin letters denote classical limiting
observables (e.g. Weyl symbols) of the corresponding operators denoted by 
capital Latin letters. 
For example, let $\ve{j}=(\ve{j}_{\rm A},\ve{j}_{\rm B})$ denote
a $d=d_{\rm A}+d_{\rm B}$ dimensional vector of classical canonical actions of the
completely integrable uncoupled classical Hamiltonian $h_0 = h_{\rm A} + h_{\rm B}$.
$d_{\rm A}$ and $d_{\rm B}$ are the number of DOF of the subsystems ``A'' and
``B'', respectively.
In quantum mechanics, one has a vector of mutually commuting 
action operators $\ve{J}$, with a common set of eigenvectors, denoted by a
multi-index $\ve{n}\in\Z^d$ of quantum numbers: $\ve{J}\ket{\ve{n}} = 
\hbar(\ve{n}+\bm{\alpha})\ket{\ve{n}}\approx \hbar\ve{n}\ket{\ve{n}}$
 where $\bm{\alpha}$ are the Maslov indices.
Here and below ``$\approx$'' means {\em equal in the leading order in} $\hbar$.
The purity (\ref{eq:IM}) can now be written as a sum over $d-$dimensional 
lattice of quantum numbers, using the fact since $\bar{V}$ commutes with $H_0$
it is diagonal in the basis $\ket{\ve{n}}$, 
and in the leading semiclassical order (in $\hbar$) we can replace the summation by an integral over the classical action space. 
Further, we replace the operator 
$\bar{V}$ by its classical limit $\bar{v}(\ve{j})$, which is a
conserved quantity so it is a function of $d$ classical actions 
$\ve{j}$ only. Let us denote by 
$p(\ve{j})=p_{\rm A}(\ve{j}_{\rm A}) p_{\rm B}(\ve{j}_{\rm B})$ the classical limit of 
the initial density $\bracket{\ve{n}}{\rho(0)}{\ve{n}}$. For our initial product state 
of two wave-packets each of the two densities is a Gaussian
\begin{equation}
p_{\rm a}(\ve{j}_{\rm a})= C \exp{\{-(\ve{j}_{\rm a}-\ve{j}_{\rm a}^*)\Lambda_{\rm a} (\ve{j}_{\rm a}-\ve{j}_{\rm a}^*)/\hbar\}},
\end{equation} 
where a subscript ``a'' takes values ``A'' or ``B'', depending on the 
subsystem, $\ve{j}_{\rm a}^*$ is the position of the initial packet, 
$\Lambda_{\rm a}$ is a positive squeezing matrix and $C=(\hbar/\pi)^{d_{\rm a}/2}\sqrt{\det{\Lambda_{\rm a}}}$ is normalization constant. 
The purity can now be written as an integral
\begin{eqnarray}
I(t)\approx \hbar^{-2d} \int{\!\! d\ve{j}\, d\tilde{\ve{j}}\, \exp{\left(-\ii \frac{\delta t}{\hbar} \Phi\right)} p(\ve{j}) p(\tilde{\ve{j}})}, \nonumber \\
\Phi=\bar{v}(\ve{j}_{\rm A},\ve{j}_{\rm B})-\bar{v}(\tilde{\ve{j}}_{\rm A},\ve{j}_{\rm B})+\bar{v}(\tilde{\ve{j}}_{\rm A},\tilde{\ve{j}}_{\rm B})-\bar{v}(\ve{j}_{\rm A},\tilde{\ve{j}}_{\rm B}).
\label{eq:Iint}
\end{eqnarray}
Next we expand the phase $\Phi$ around the position 
$\ve{j}^*=(\ve{j}_{\rm A}^*,\ve{j}_{\rm B}^*)$ of the initial packet. 
The constant and the linear terms cancel exactly and the lowest order non-vanishing 
term is quadratic
\begin{equation}
\Phi \approx (\ve{j}_{\rm A}-\tilde{\ve{j}}_{\rm A})\cdot \bar{v}''_{\rm AB}(\ve{j}^*) (\ve{j}_{\rm B}-\tilde{\ve{j}}_{\rm B}) + \cdots,
\label{eq:Phi}
\end{equation}
where $\bar{v}''_{\rm AB}$ is a $d_{\rm A} \times d_{\rm B}$ matrix 
of mixed second derivatives of $\bar{v}$ evaluated at the position of the initial 
packet,
\begin{equation}
\left( \bar{v}''_{\rm AB} \right)_{kl}=\frac{\partial^2 \bar{v}}{\partial (\ve{j}_{\rm A})_k \partial(\ve{j}_{\rm B})_l}.
\label{eq:vce}
\end{equation}
Using this expansion in the integral for purity we see that the resulting $2d$ dimensional integral is Gaussian and can therefore be expressed in terms of a determinant of a 
$2d\times 2d$ matrix. Using special properties of the resulting matrix the determinant 
can be reduced~\cite{Znidaric:04} to a determinant of a $d_{\rm A}\times d_{\rm A}$
 matrix, with the final result
\begin{equation}
I(t)=\frac{1}{\sqrt{\det{(\mathbbm{1}+(\delta t)^2 u)}}},\quad u=\Lambda_{\rm A}^{-1} \bar{v}''_{\rm AB} \Lambda_{\rm B}^{-1} \bar{v}''_{\rm BA},
\label{eq:Imain}
\end{equation}
where $u$ is a $d_{\rm A}\times d_{\rm A}$ matrix involving 
$\bar{v}''_{\rm AB}$ and its transpose $\bar{v}''_{\rm BA}$. Note that the 
matrix $u$ is a classical quantity (independent of $\hbar$) that depends 
only on observable $\bar{v}$ and on the position of the initial packet. 
This explicit formula for purity decay is the main result of the 
present paper~\footnote{The very same expression holds 
also for a generalization of purity to echo dynamics, so-called echo purity (or purity fidelity), first used in~\cite{Prosen:02spin}.}.

Before discussing its consequences let us remind on its range of validity. 
The restrictions are rather weak: $\bar{v}$ must be nonvanishing 
(typical for regular systems) and smooth on a scale of the initial packet
$\propto \sqrt{\hbar}$, time must be larger than the averaging time 
$t> t_{\rm ave}$, and the coupling must be small $\delta<1$. In addition, the phase $\Phi$ 
should increment by a small amount for neighboring quantum numbers, which 
translates into the condition $\delta t ||\bar{v}''_{\rm AB}|| < 1/\hbar$.

The most prominent feature of the formula (\ref{eq:Imain}) for the purity 
decay for initial product wave packets is its $\hbar$ independence. 
In the linear response calculation this $\hbar$-independence has already 
been theoretically predicted~\cite{Prosen:03evol} as well as numerically 
confirmed~\cite{Znidaric:03}. Here we have a full expression to all orders. 
We also see that the scaling of the decay time $t_{\rm d}$ on which $I(t)$ 
decays is $t_{\rm d} \sim 1/\delta$. This means that the purity will 
decay on a very long time scale and so the wave packets are universal 
pointer states~\cite{Zurek:91}, i.e. the most robust states. 
For small $\delta t$ we can 
expand the determinant and we get initial quadratic decay $I(t)=1-\frac{1}{2}(\delta t)^2 \tr{}{u}+\cdots$. For large times we use the fact that $\det(\mathbbm{1} + z u)$ is
a polynomial in $z$ of order $r={\rm rank\,}(u)$, so we have asymptotic power law decay
$I(t) \asymp {\rm const}\,(\delta t)^{-r}$.
Note that the rank of $u$ is bounded by the minimal of the subspace dimensions,
i.e. 
$1\le r \le {\rm min}\{d_{\rm A},d_{\rm B}\}$, 
since the definition (\ref{eq:I}) is symmetric with respect to 
interchanging the roles of the subspaces ``A'' and ``B''.
Let us give two simple examples:
(i) For $d_{\rm A}=1$ and for {\em any} $d_{\rm B}$ we will always 
have asymptotic power law decay with $r=1$. If a {\em single} DOF of 
the subsystem ``A''  is coupled with {\em all} DOF of the subsystem ``B'', e.g. 
$\bar{v}=j_{\rm A} \otimes (j_{\rm B1}+j_{\rm B2}+\cdots)$, then 
$|\bar{v}''|^2 \propto d_{\rm B}$ and we have 
$I(t) \asymp 1/(\delta t \sqrt{d_{\rm B}})$; (ii) Let us 
consider a multidimensional system where the matrix $u$ is of rank one so it
can be written as a direct product of 
two vectors, $u=\ve{x} \otimes \ve{y}$. The determinant occurring in $I(t)$ 
is then simply 
$\det{(\mathbbm{1}+(\delta t)^2u)}=1+(\delta t)^2 \ve{x}\cdot \ve{y}$. 
Such is the case for instance if we have a coupling of the same strength 
between all pairs of DOF. The dot product is in this case 
$\ve{x}\cdot \ve{y} \propto d_{\rm A} d_{\rm B}$ and we have 
$I(t) \asymp 1/(\delta t \sqrt{d_{\rm A} d_{\rm B}})$, i.e. the power of 
the algebraic decay is independent of both $d_{\rm A}$ and $d_{\rm B}$. 

In~\cite{Jacquod:04} the author predicted a universal asymptotic 
$t^{-2}$ decay of purity independent of the dimensions or the coupling involved. He obtained this result for an average purity, i.e. averaged over the position of the initial 
packet. Our result for $I(t)$ (\ref{eq:Imain}) clearly can not reproduce 
the result of Ref.~\cite{Jacquod:04}, even if we average over the position of 
the initial packet. 
Such an average decay will in general depend on the functional 
dependence of the matrix $u$ on the position of the initial packet. 

%As for the 
%numerical results of Fig.~2 of~\cite{Znidaric:03}, which the authors claims 
%to explain as being $t^{-2}$, we must add that the highly oscillatory 
%decay seen in that figure can not be associated with a well defined power law. 
%This non generic result was a consequence of a resonance condition betwen 
%both subsystems. In a general case one gets a $t^{-1}$ decay as predicted by our 
%theory, see Fig.~5.4 in~\cite{Znidaric:04}. 

\begin{figure}
\centerline{\includegraphics[height=3.3in,angle=-90]{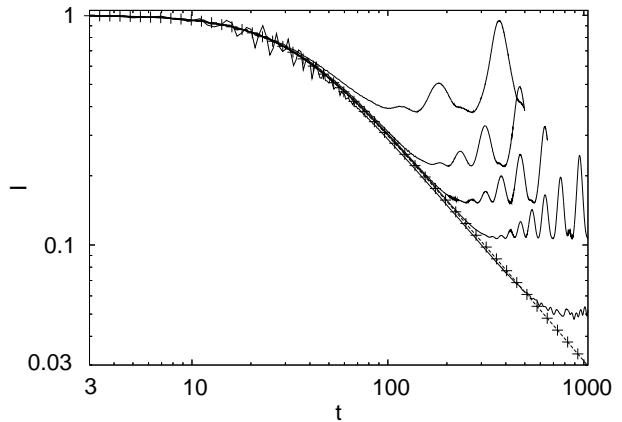}}
\caption{Purity decay for $1 + 1$ DOF system (\ref{eq:H01x1},\ref{eq:V1x1}) 
for $\delta=0.04$ and different $1/\hbar=10,25,50,100,500$, from top to 
bottom. Dashed line with pluses is the theoretical formula (\ref{eq:I1x1}).}
\label{fig:1x1hbar}
\end{figure}

\begin{figure}
\centerline{\includegraphics[height=3.3in,angle=-90]{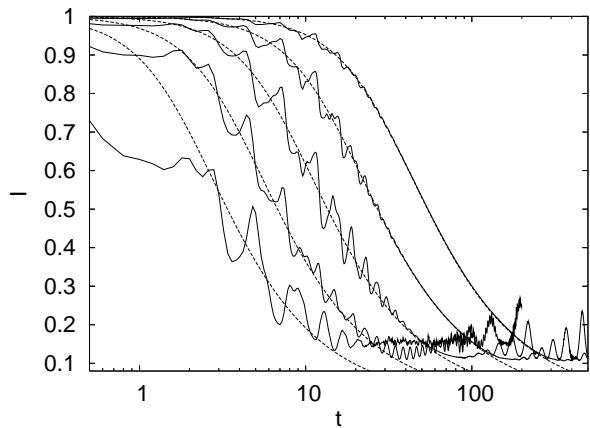}}
\caption{Purity decay for $1 + 1$ DOF system (\ref{eq:H01x1},\ref{eq:V1x1}) 
for $\hbar=1/100$ and different $\delta=0.64, 0.32, 0.16, 0.08, 0.04$, 
from left to right. Dashed lines give the theoretical prediction (\ref{eq:I1x1}).}
\label{fig:1x1delta}
\end{figure}

We continue with a numerical demonstration of the theoretical prediction for purity decay (\ref{eq:Imain}). For the first example we take a $1 + 1$ DOF system, 
$d_{\rm A}=d_{\rm B}=1$, of two an-harmonic oscillators with the uncoupled Hamiltonian 
\begin{equation}
H_0= \gamma_{\rm A} (\hbar a_{\rm A}^+ a_{\rm A}-\Delta)^2+ \gamma_{\rm B} (\hbar a_{\rm B}^+ a_{\rm B}-\Delta)^2,
\label{eq:H01x1}
\end{equation}
where $a^+,a$ are standard boson raising/lowering operators. For the 
coupling we take
\begin{equation}
V=\hbar^2 (a_{\rm A}^+ + a_{\rm A})^2(a_{\rm B}^+ + a_{\rm B})^2.
\label{eq:V1x1}
\end{equation}
The corresponding classical Hamiltonian $h$ reads
\begin{equation}
h=\gamma_{\rm A} (j_{\rm A}-\Delta)^2+\gamma_{\rm B} (j_{\rm B}-\Delta)^2 + 
16 \delta j_{\rm A} j_{\rm B} \sin^2{\theta_{\rm A}} \sin^2{\theta_{\rm B}}
\end{equation} 
where $\theta_{\rm a}$ are the canonical angles.
The initial wave packet on both subsystems is a boson coherent state
\begin{equation}
\ket{\psi_{\rm A}(0)}=\ket{\psi_{\rm B}(0)}=\ket{\alpha}={\rm e}^{\alpha a^+ - \alpha^* a} \ket{0},
\label{eq:boson_coh}
\end{equation}
where $\ket{0}$ is the ground state. The parameter $\alpha$ is chosen as $\alpha=\sqrt{j^*/\hbar}$ with $j^*=0.1$. The squeezing parameter for the coherent states 
(\ref{eq:boson_coh}) is $\Lambda_{\rm A,B}=1/(2j^*)$. Other parameters of the 
Hamiltonian are $\gamma_{\rm A}=1$, $\gamma_{\rm B}=0.6456$. The offset 
$\Delta=1.2$ was chosen in order to have nonzero classical frequency 
$\partial h/\partial j$ 
at the position of the initial packet. This is needed in order for $\bar{v}$ to be 
well defined. Time averaged coupling is calculated easily,
$\bar{v}=4 j_{\rm A} j_{\rm B}$.
The matrix $u$ is now just a number, $u=(8j^*)^2$. Theoretical 
prediction for the purity decay is thus
\begin{equation}
I(t)=\frac{1}{\sqrt{1+(8 j^* \delta t)^2}}.
\label{eq:I1x1}
\end{equation}
The results of numerical simulation together with the theory are shown in Figs.~\ref{fig:1x1hbar} and~\ref{fig:1x1delta}.
In Fig.~\ref{fig:1x1hbar} we see that the decay is indeed $\hbar$-independent, 
apart from a finite size fluctuating plateau after long time. The size of this 
plateau is of the order $I(t \to \infty) \sim 1/N_{\rm eff}$,
where $N_{\rm eff}\sim \sqrt{8j^*/\hbar}$ is an effective Hilbert space 
dimension, i.e. the number of action eigenstates overlaping with the 
initial coherent state (\ref{eq:boson_coh}). Strong revivals for large $\hbar$ are a consequence of small number of 
available states $N_{\rm eff}$ and low dimensionality. Revivals are expected 
to be less pronounced for larger dimensionalities $d_{\rm A},d_{\rm B}$, similarly as 
for the fidelity~\cite{Prosen:02JPA}. For large times one can clearly observe 
asymptotic $t^{-1}$ decay of the purity.
In Fig.~\ref{fig:1x1delta} we fix $\hbar$ and change the coupling 
strength $\delta$ instead. Apart from oscillations we see a good 
agreement with the theory also for large $\delta$. Oscillations for 
times $t<10$ are a consequence of the fact that the time averaging of 
$V$ (\ref{eq:Vbar}) converges only after some averaging time $t_{\rm ave}$ 
which is of order $\sim 10$ in our case.

\begin{figure}[h]
\centerline{\includegraphics[angle=-90,width=3.3in]{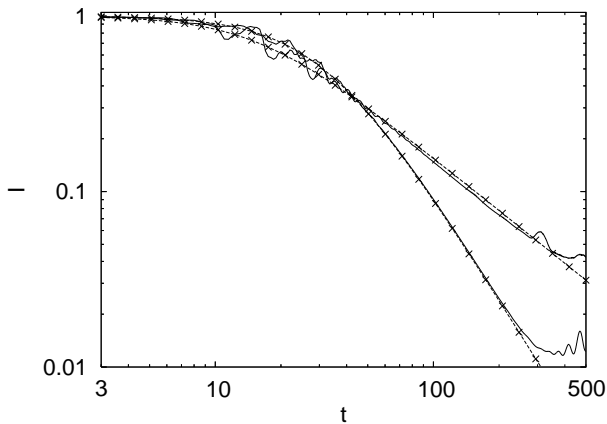}}
\caption{Purity decay for a $2 + 2$ DOF system (\ref{eq:H02x2}) and two 
different couplings showing different asymptotic power law decay. Full line 
is the numerics while the two dashed lines with crosses are theoretical 
predictions, the one with a smaller slope for (\ref{eq:I2x2_B}) and the 
other for (\ref{eq:I2x2_A}).}
\label{fig:2x2}
\end{figure}
As for the second numerical example we take a $2 + 2$ DOF system ($d_{\rm A,B}=2$) 
which is the simplest case where we can find different power of the asymptotic 
decay, depending on the topology of the coupling. The uncoupled Hamiltonian now reads
\begin{eqnarray}
H_0 &=& \gamma_1 (\hbar a_1^+ a_1-\Delta)(\hbar a_2^+ a_2-\Delta) + \nonumber \\
&+& \gamma_2 (\hbar a_3^+ a_3 -\Delta)(\hbar a_4^+ a_4 -\Delta).
\label{eq:H02x2}
\end{eqnarray}
Subscripts ``1'' and ``2'' describe two DOF of the subsystem ``A'', while ``3'' and ``4'' compose the subsystem ``B''. The parameters are $\gamma_1=1$, $\gamma_2=0.64$ and $\Delta=1.2$. The initial state is a product state of four boson coherent states, $\ket{\psi_{1,2,3,4}(0)}=\ket{\alpha}$, all 
with the same $\alpha=\sqrt{j^*/\hbar}$. For the coupling we consider two 
cases which will give different power of the asymptotic decay. {\em Case I.}: 
$V=V_{13}+V_{24}$, 
where two coupling terms are of the same form as for $1 + 1$ DOF system (\ref{eq:V1x1}) and the indices denote between which two degrees of freedom the coupling acts. The matrix $u$ as well as the relevant determinant is easily calculated 
resulting in a simple expression for the purity (\ref{eq:Imain})
\begin{equation}
I(t)=\frac{1}{1+(8j^* \delta t)^2}.
\label{eq:I2x2_A}
\end{equation}  
We see that we have a quadratic asymptotic decay, $I(t) \asymp 1/(\delta t)^2$.
{\em Case II.}: All to all coupling,
$V=V_{13}+V_{14}+V_{23}+V_{24}$,
results in a rank-one ($r=1$) matrix $u$ giving the 
purity decay (\ref{eq:Imain})
\begin{equation}
I(t)=\frac{1}{\sqrt{1+(16 j^* \delta t)^2}}.
\label{eq:I2x2_B}
\end{equation}
Results of the numerical simulation for both cases are shown in Fig.~\ref{fig:2x2}. The coupling strength and the location of the initial packets are 
$\delta=0.04$, $j^*=0.1$ for the case {\em I}., 
and $\delta=0.02$, $j^*=0.2$ for the case {\em II}. 
From Fig.~\ref{fig:2x2} we see that one indeed has asymptotic 
$t^{-1}$ or $t^{-2}$ decay, depending on the topology of the coupling.

In conclusion, we have derived purity decay for initial localized wave-packets 
in bipartite systems with a non-vanishing (non-trivial) time-averaged coupling operator.
Such situation naturally occurs in systems where an uncoupled part of the Hamiltonian 
represents regular dynamics. Purity decays in time inversely proportional to the 
coupling strength and is independent of Planck's constant. The decay is 
algebraic with the asymptotic power-law exponent ranging between $1$ and 
the minimal dimension of the
subsystems depending on the topology of the coupling.

We thank Thomas H. Seligman for fruitful discussions. 
Financial supports by the grant P1-044 of the Ministry of Education, Science and 
Sports of Slovenia, and in part by the ARO grant (USA) DAAD 19-02-1-0086,
and hospitality of CiC (Cuernavaca, Mexico), 
where parts of this work have been completed, are gratefully acknowledged.

\end{document}